# Microstructure of a spark-plasma-sintered $Fe_2VAl$-type Heusler alloy for thermoelectric application

*Leonie Gomell[a], Imants. Dirba[b], Hanna Bishara[a,1], Zhongji Sun[a,2], Łukasz. Żrodowski[e,f], Tomasz Choma[e,f], Bartosz Morończyk[e,f], Gerhard Dehm[a], Konstantin P. Skokov[b], Oliver Gutfleisch[b], Baptiste Gault[a,g,]\**

[a] Max-Planck-Institut für Eisenforschung GmbH, Max-Planck-Str. 1, 40237 Düsseldorf, Germany

[b] Institute of Materials Science, Technische Universität Darmstadt, 64287 Darmstadt, Germany

[e] AMAZEMET sp. z o.o,. al. Jana Pawla II 27, Warsaw Poland

[f] Warsaw University of Technology Faculty of Materials Science and Engineering, Woloska 141, Warsaw Poland

[g] Department of Materials, Royal School of Mines, Imperial College London, London, UK

\* Corresponding author: b.gault@mpie.de

Present address:

1) Materials Science and Engineering, Tel Aviv University, P.O. Box 39040, Tel Aviv 6997801, Israel
2) Institute of Materials Research and Engineering, 2 Fusionopolis Way, Singapore 138634

Email addresses:

L. Gomell:        l.gomell@mpie.de

I. Dirba:         imants.dirba@tu-darmstadt.de

H. Bishara:       hbishara@tauex.tau.ac.il

Z. Sun:           sun_zhongji@imre.a-star.edu.sg

Ł. Żrodowski:     lukasz.zrodowski@amazemet.com

T. Choma:         tomasz.choma@amazemet.com

B. Morończyk:     bartosz.moronczyk@amazemet.com

G. Dehm:          dehm@mpie.de

K.P. Skokov:      konstantin.skokov@tu-darmstadt.de

O. Gutfleisch:    oliver.gutfleisch@tu-darmstadt.de

B. Gault:         b.gault@mpie.de

# Abstract

The influence of microstructure on thermoelectricity is increasingly recognized. Approaches for microstructural engineering can hence be exploited to enhance thermoelectric performance, particularly through manipulating crystalline defects, their structure, and composition. Here, we focus on a full-Heusler $Fe_2VAl$-based compound that is one of the most promising thermoelectric materials containing only Earth-abundant, non-toxic elements. A $Fe_2VTa_{0.05}Al_{0.95}$ cast alloy was atomized under a nitrogen-rich atmosphere to induce nitride precipitation. Nanometer- to micrometer-scale microstructural investigations by advanced scanning electron microscopy and atom probe tomography (APT) are performed on the powder first and then on the material consolidated by spark-plasma sintering for an increasing time. APT reveals an unexpected pick-up of additional impurities from atomization, namely W and Mo. The microstructure is then correlated with local and global measurements of the thermoelectric properties. At grain boundaries, segregation and precipitation locally reduce the electrical resistivity, as evidenced by *in-situ* four-point probe measurements. The final microstructure contains a hierarchy of structural defects, including individual point defects, dislocations, grain boundaries, and precipitates, that allow for a strong decrease in thermal conductivity. In combination, these effects provide an appreciable increase in thermoelectric performance.

# 1 Introduction

Thermoelectric (TE) materials generate electricity from (waste) heat, and can hence, contribute to more sustainable energy production. Designing efficient TE materials employing earth-abundant and cost-efficient elements remains a challenge. $Fe_2VAl$-based materials fulfill the second requirement [1], yet, their efficiency is insufficient to compete with state-of-the-art materials [2,3]. The efficiency is usually described by the figure of merit $zT = \frac{S^2}{\kappa\rho}T$, where *S* is the Seebeck coefficient, $\rho$ is the electrical resistivity, $\kappa$ is the thermal conductivity, and *T* is the temperature [4]. $Fe_2VAl$ shows an intrinsically high power factor $S^2/\rho$ of approximately 5 mW m$^{-1}$ K$^{-2}$, which can compete with e.g. $Bi_2Te_3$ [1]. However, the thermal conductivity is an order of magnitude higher, thereby limiting the efficiency.

Microalloying is a known design path to reduce the thermal conductivity, increase the electrical conductivity and the Seebeck coefficient, and hence optimize TE performance [5–7]. In addition, a hierarchical microstructure with defects on all length scales – from point defects to dislocations and grain boundaries (GBs) to the second-phase precipitates – scatters phonons of different wavelengths leading to a decrease in the thermal conductivity [8]. For $Fe_2VAl$, we recently showed that microstructural manipulations by laser surface remelting, ensuring a fast solidification rate, increased the TE performance [9,10], and that microalloying via *in-situ* nitriding enabled us to control the composition of the Cottrell atmosphere around dislocations and GBs, leading to an enhancement of the thermoelectric properties [11]. In bulk samples, Ta-doping was reported to increase *zT* by an order of magnitude from 0.01 for undoped, cast $Fe_2VAl$ to 0.2 for $Fe_2VTa_{0.05}Al_{0.95}$ [1,12]. Combining optimized composition and a microstructure with e.g. small grain size, high dislocation density etc. is an appealing route to further enhance TE performance.

Here, we report on the synthesis by atomization of powder with a nominal composition of $Fe_2VTa_{0.05}Al_{0.95}$, under a N-rich atmosphere. Fast sintering by spark plasma sintering (SPS) [13] was then used to obtain a dense bulk material. SPS leads to minimized grain growth [14], which helps to achieve a high density of phonon scattering centers. By interrupting the SPS process for times ranging from 15 seconds to 5 minutes, we monitor the development of the microstructure by a combination of scanning electron microscopy (SEM), including backscattered electron (BSE) imaging, electron-backscattered diffraction (EBSD) and electron-channeling contrast imaging (ECCI), and atom probe tomography (APT). These measurements are correlated to local and bulk measurements of transport properties to discuss how the TE performance is affected by the array of chemically decorated microstructural features, including dislocations and faults, and secondary phases, depending on the composition of each powder grain. We reveal an unexpected contamination during powder atomization, likely to affect other materials processed using similar path, which, in our case may further help enhance properties. Our combined results show how this material proves to be tolerant to impurities, particularly through our proposed processing route, opening a possibility for the future production of TE generators without the need of using costly and toxic elements.

## 2   Experimental

Pure Fe (99.9%, Carboleg GmbH), V, Ta (both 99.9%, HMW Hauner GmbH), and Al (99.7%, Aluminium Norf GmbH) were used to produce powder with a nominal composition of $Fe_2VTa_{0.05}Al_{0.95}$ by ultrasonic atomization in nitrogen atmosphere. Vanadium reacts with the nitrogen atmosphere to vanadium nitrides at high temperatures, introducing segregation of V and N to defects [11]. The ultrasonic atomization and nitriding were realized with the rePowder device (AMAZEMET, Poland) in plasma mode [19]. The nitriding process was carried out at an absolute pressure of 1400 mbar Ar atmosphere and 10 mbar $N_2$. A molybdenum sonotrode was used at 40 kHz frequency and 30 µm amplitude. Plasma current was kept at 150-200 A during the process.

Representative powder particles were embedded and polished down to 0.05 µm colloidal silica to obtain information on the cross section of the particles. The rest of the powder was consolidated: 1 g powder was sintered using SPS in an 8 mm diameter graphite die. The temperature was optimized to be 1000 °C, as lower temperatures lead to poor mechanical integrity and higher temperatures lead to partial melting, which should be avoided. A pressure of 60 MPa was applied during the sintering process. The holding time was chosen to be 15 seconds, 30 seconds, 1 minute, 2 minutes, and 5 minutes to understand the microstructural changes during SPS. To distinguish between the different samples, the holding time will be given as a subscript. The disks were cut centrally to expose the cross-section and polished down to 0.05 µm colloidal silica before an investigation of the microstructure.

Scanning electron microscopy (SEM), including electron backscattered diffraction (EBSD), and energy-dispersive X-ray spectroscopy (EDX), were performed on all samples using a Zeiss Sigma. The acceleration voltage was set to 20 kV for BSE imaging, EDX, and EBSD. EBSD mapping was conducted with a step size of 250 nm on a hexagonal grid.

The as-produced powder and the $SPS_{5\ min}$ sample were investigated more closely by electron channeling contrast imaging (ECCI) and atom probe tomography (APT). ECCI was conducted using a Zeiss Merlin with an acceleration voltage of 30 kV. APT specimens were prepared using a dual-beam focused ion beam instrument (FEI Helios Nanolab 600i), equipped with a $Ga^+$ ion source, as described in Ref. [36]. APT was conducted using a LEAP 5000 XS instrument (Cameca Instruments) operated in laser pulsing mode. The pulse energy was set to 50 pJ, the pulse repetition rate to 200 kHz, and the detection rate to 2 %. The base temperature was kept at 60 K. The data was reconstructed and analyzed by AP Suite 6.1 (Cameca Instruments).

For the SPS$_{5\,min}$ sample, simultaneously measurements of the sample's thermal conductivity, Seebeck coefficient, and electrical resistivity were carried out using the Thermal Transport Option for PPMS14 system in temperature range 16 K-400 K. To understand the impact of the different defects, i.e. GBs, and different precipitates on the electrical resistivity, we use a local in-situ four-point probe technique inside a Zeiss Auriga SEM. In this system, four probes with a tip radius of 50 nm can be positioned by independent micromanipulators (PS4, Kleindiek Nanotechnik GmbH). The mutual distance between the probes was approximately 5 µm. The resistivity value is calculated based on resistance and distance between needles. The error within the resistance is set as a standard deviation of repeating 100 measurements [31]. In addition, each point is the average of 3–4 measurements in the same grain or at the same GB. The deviation of the distance between needles is set to 50 nm, estimated by the SEM resolution and tip radius. A description of errors, corrections due to deviation from equidistance positions of needles, and modulation of electrical current to decrease the noise of this measurement is given in Ref. [31].

## 3 Results and Discussion

### 3.1 Microstructure of the powder

The as-produced powder (see Methods), **figure S1**a-b, consists of spherical particles with a mean diameter of (54 ± 8) µm. BSE imaging of the cross-section of a powder particle, **figure 1**a, shows a dendritic structure with brightly imaged interdendritic regions, indicating the presence of heavier elements (i.e. Z-contrast). EDX evidences an enrichment in Ta and a V depletion, **figure S2**. Such an inhomogeneous distribution of elements is typically observed in powder from atomization of compositionally complex alloys [15–17]. We observed 150 individual particles, and approx. 99% of particles show this dendritic structure, either as monocrystalline or bicrystalline particles. The remaining 1% are polycrystalline particles with no visible dendritic structure. We can assume that these particles re-entered the hot flame during the atomization, leading to remelting or simply a heat treatment.

Only 26% of all particles show a composition by EDX close to the nominal composition. In approx. 66% of the particles, Mo is found, mostly in minor levels <2 at.%, but in rare occasions also up to 30 at.% in a single particle. In addition, W is found in approx. 12% of the particles. These impurities originate from the baseplate of the atomizer. The ingress of impurities from the base plate is likely a common issue in powder atomization, as previously reported [18], including for the specific crucible used here that showed systematic ingress of W for multiple powders [19]. Here, both W and Mo are n-type dopants in the full-Heusler Fe$_2$VAl, which can lead to a positive effect on the TE properties [24,25]. However, these unintended compositional changes from the processing make the optimization of the charge carrier concentration more difficult. The impurity concentrations are partly accompanied by depletion of Al and enrichment of V. No correlation between microstructure and composition was observed, **figure S3**. **Table S1** gives further information about correlations between enrichment and depletion of different elements within individual particles.

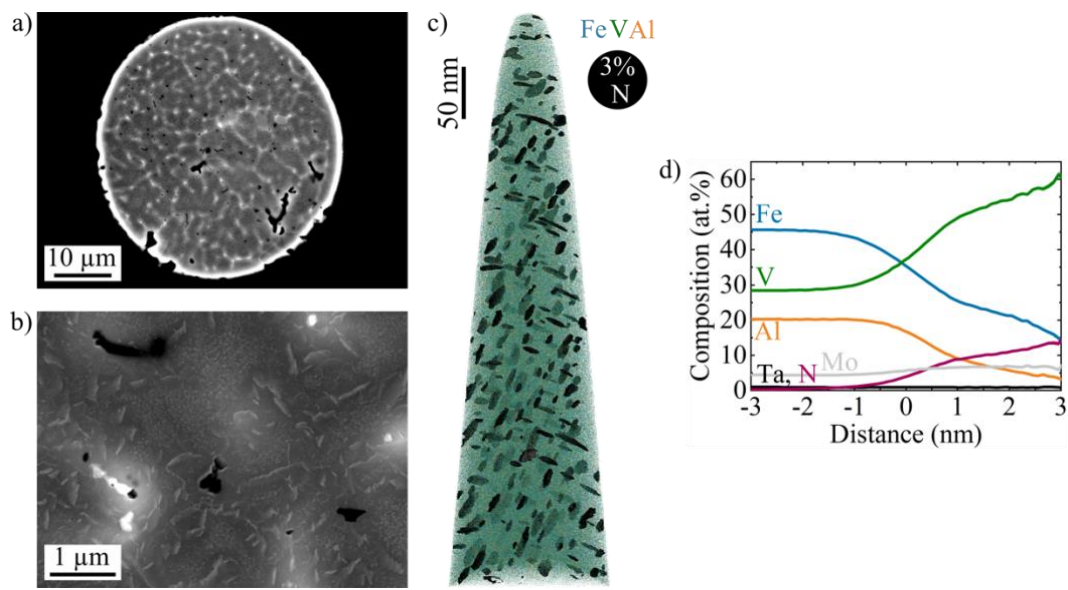

*Fig. 1*: *a) Backscattered electron micrographs of a $Fe_2VTa_{0.05}Al_{0.95}$ powder particle, which shows a dendritic structure with bright imaged interdendritic regions. b) Close-up of the microstructure, showing a high density of precipitates. c) APT reconstruction, showing a high density of disc-shaped $VN_x$ precipitates, indicated by a black iso-composition surface of 3 at.% N. Fe, V, and Al are shown as blue, green, and orange dots, and d) proxigram of the precipitates.*

High-magnification BSE images reveal platelet-shaped precipitates with a density of approx. $10^{14}$ m$^{-2}$ (Figure 1b and S1d). The precipitates are approx. 50–200 nm in length, along with a second population of larger particles approx. 1µm in length. APT. An APT analysis from the center of a particle with a dendritic structure is displayed in figure 1c, and precipitates are highlighted by black iso-composition surfaces delineating regions with a composition threshold of 3 at.% N. The precipitates grew on the (001) plane of the $Fe_2VAl$ matrix, as indicated by the 90° angle between different precipitates. A composition profile calculated as a function of the distance to the iso-surface (proximity histogram or proxigram [26]), figure 1d, reveals enrichment of V and N, Fe and Al depletion, pointing to $VN_x$ precipitates previously reported [10,27]. The Ta composition is the same in and out of the precipitates. In addition, we measure approx. 5 at.% Mo. The composition within the matrix and inside the precipitates is given in **table 1**. W was also found in other APT specimens in the range of a few atomic %.

*Table 1: Composition of the matrix and the precipitates, determined by APT.*

|  | Fe [at.%] | V [at.%] | Al [at.%] | Ta [at.%] | N [at.%] | Mo [at.%] |
|---|---|---|---|---|---|---|
| **Matrix** | 45.7 ± 0.1 | 28.3 ± 0.1 | 20.3 ± 0.1 | 1.1 ± 0.1 | 0.2 ± 0.1 | 4.4 ± 0.1 |
| **precipitate** | 14.6 ± 2.1 | 59.8 ± 2.1 | 3.7 ± 0.8 | 1.0 ± 0.2 | 14.1 ± 1.4 | 6.8 ± 1.2 |

### 3.2 Microstructure of the SPS-consolidated materials

The powder was consolidated into a bulk material by SPS, in which the powder particles are pressed, but remain as individual grains in the bulk material. Depending on the sintering time, we observe a partly-to-fully-dense disc, as shown in the cross-sectional BSE images in **figure 2** a–e. Samples are denoted SPS$_t$ with $t$ ranging from 15s to 5 mins. The density, estimated from the electron micrographs, figure 2f, increases from approx. 94.0 % to 99.5 % respectively for SPS$_{15s}$ and SPS$_{5min}$. The dendritic structure of the powder particles is no longer observed after SPS. Instead, a distribution of bright and dark precipitates (Z-contrast) is observed in all samples, with precipitates primarily concentrated in some grains.

The differences in the precipitation behavior of the various elements in the EDX maps in **figure 3**a for the SPS$_{5min}$ sample, allow us to speculate that grains with dense precipitation originated from powder particles with a higher concentration of the precipitate-forming elements. For example, approx. 1 µm Ta-rich and 150 nm W-rich precipitates are distributed across individual, separate grains. Mo does not appear precipitated on the scale imaged by SEM, but as diffuse segregation inside some grains. Al segregates to GBs, where it sometimes forms larger precipitates. Indications of slight precipitate growth for longer sintering times are noticed, as analyzed for Ta precipitates, **figure S4**, yet no quantitative analysis was achievable here due to a lack of appropriate statistic.

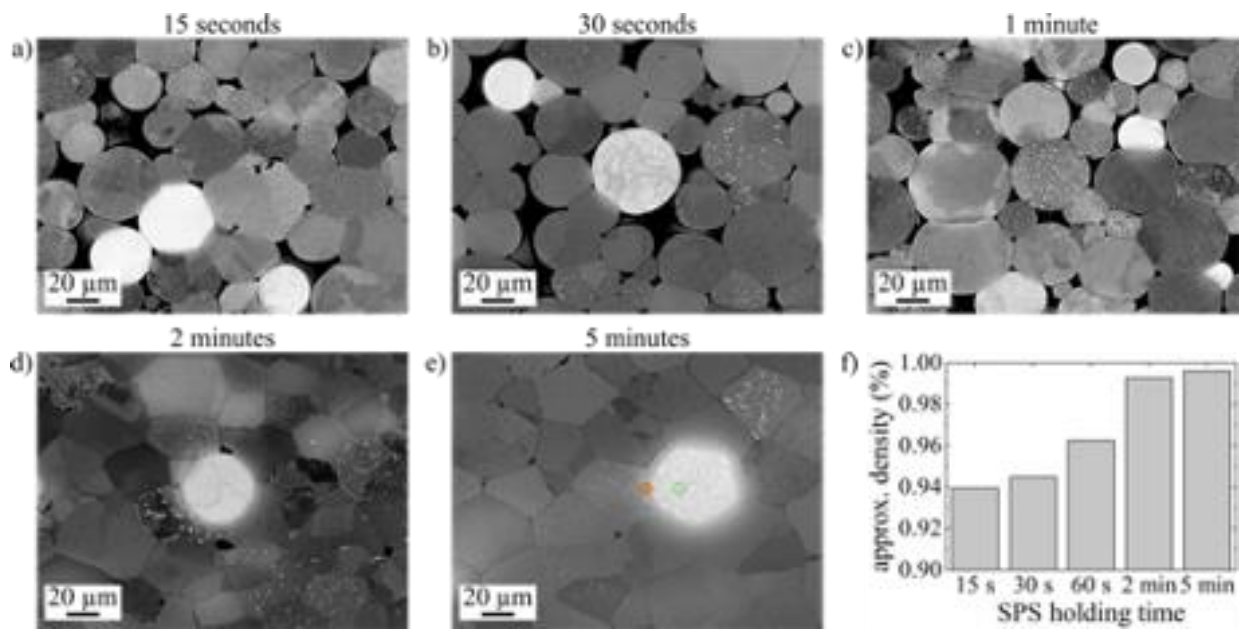

*Fig. 2* Backscattered electron images of the cross-section after SPS for different holding times: a) 15 seconds, b) 30 seconds, c) 1 minute, d) 2 minutes, and e) 5 minutes. f) Estimated relative density.

For SPS$_{5min}$, the EDX maps are complemented by an EBSD inverse pole figure and a kernel average misorientation map (KAM), **figure 3**b. The KAM shows the misorientation between two adjacent measurement positions, given on a scale from 0° to 2°, highlighting the presence of numerous low-angle GBs (LAGBs). Grains containing more Mo or W appear more rounded and with fewer LAGBs. During SPS, the powder is compacted by deformation and particle rearrangement [28,29]. The inhomogeneity of deformed particles and undeformed, spherical particles can be interpreted as arising from the inhomogeneous chemical distribution across powder particles before SPS that leads to a range of hardness, which, after SPS results in this microstructure.

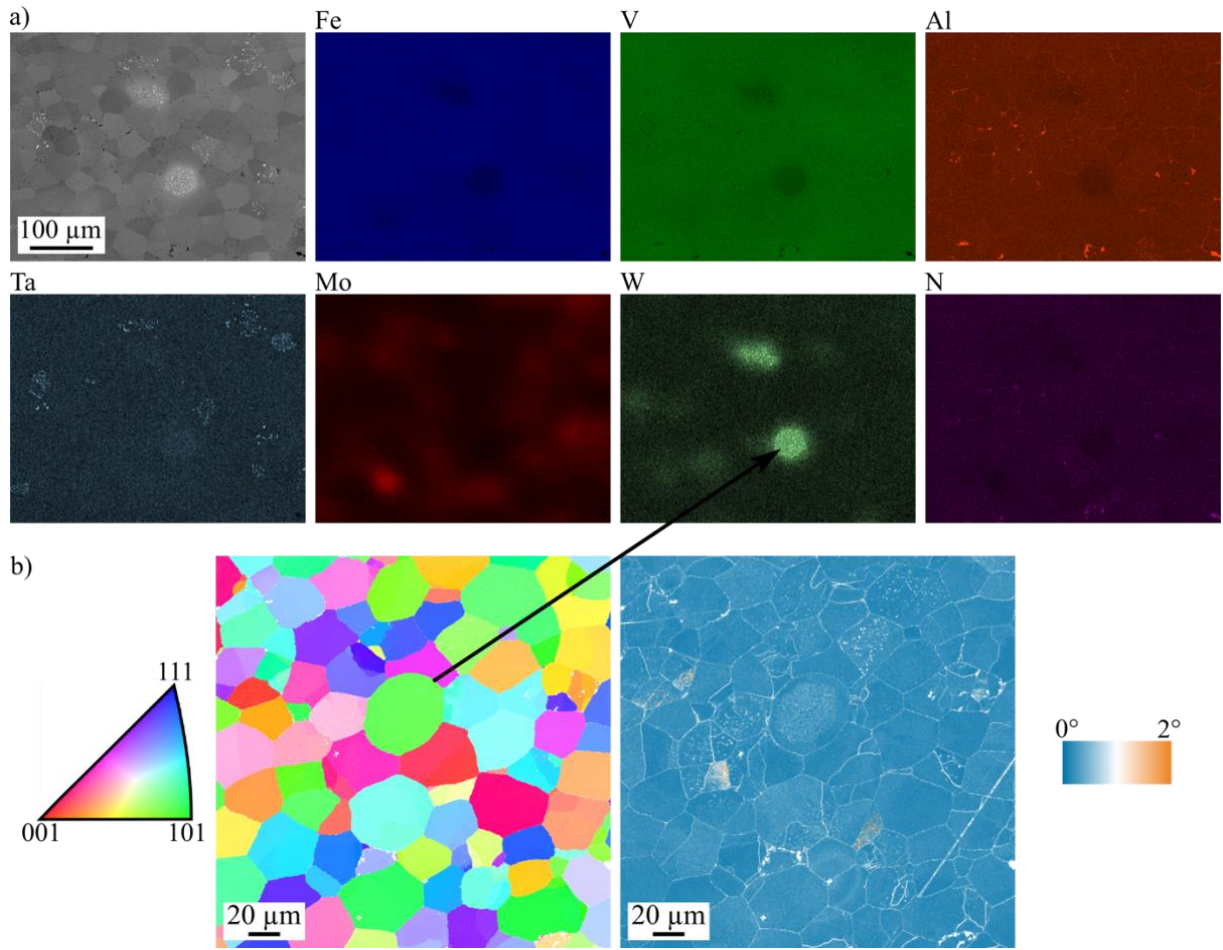

*Fig. 3 a) EDX maps of the SPS$_{5\ min}$ sample, indicating higher W and Mo content in some grains and reflecting most likely differences in precipitation. b) EBSD investigations. Left: inverse pole figure in normal direction, right: Kernel average map showing misorientation of 0°-2°. The arrow connects the same grain in the EDX and EBSD map for better orientation.*

APT was performed in the core of precipitate-rich grains and close to GBs. Equivalent representative regions are indicated by the orange and green circles in figure 2. **Figure 4**a–b shows two reconstructions and the corresponding proxigrams of the imaged precipitates. Close to a high-angle GB, in figure 4a, are Al-rich precipitates with a composition of approx. $Al_{82.1\pm0.9}Fe_{14.0\pm0.8}N_{2.5\pm0.2}$ (at.%). This composition agrees well with the solubility of Fe in the Al liquid phase at 1000 °C [30], which is the temperature during SPS. This APT analysis also shows larger Mo-rich precipitates highlighted by red isosurfaces (20 at.% Mo threshold), and platelet-shaped precipitates, probably growing at LAGBs shown by pink iso-composition (10 at.% Mo threshold). Another APT analysis from a W-rich grain shows dislocations with a W-rich Cottrell atmosphere and a W-rich precipitate, with a composition $W_{66.7\pm1.4}V_{19.2\pm1.2}Mo_{5.4\pm0.5}Ta_{5.1\pm0.4}Fe_{2.4\pm0.3}$ (at.%). Figure 4c–e are a series of matching defects imaged by ECCI and APT, including entangled dislocations, most likely forming a LAGB, from which plate-shaped precipitates form in figure 4c; a combination of partial dislocations Cottrell atmospheres in figure 4d; and a LAGB consisting of parallel dislocations, figure 4e. These examples underline the complexity of the microstructure and its composition.

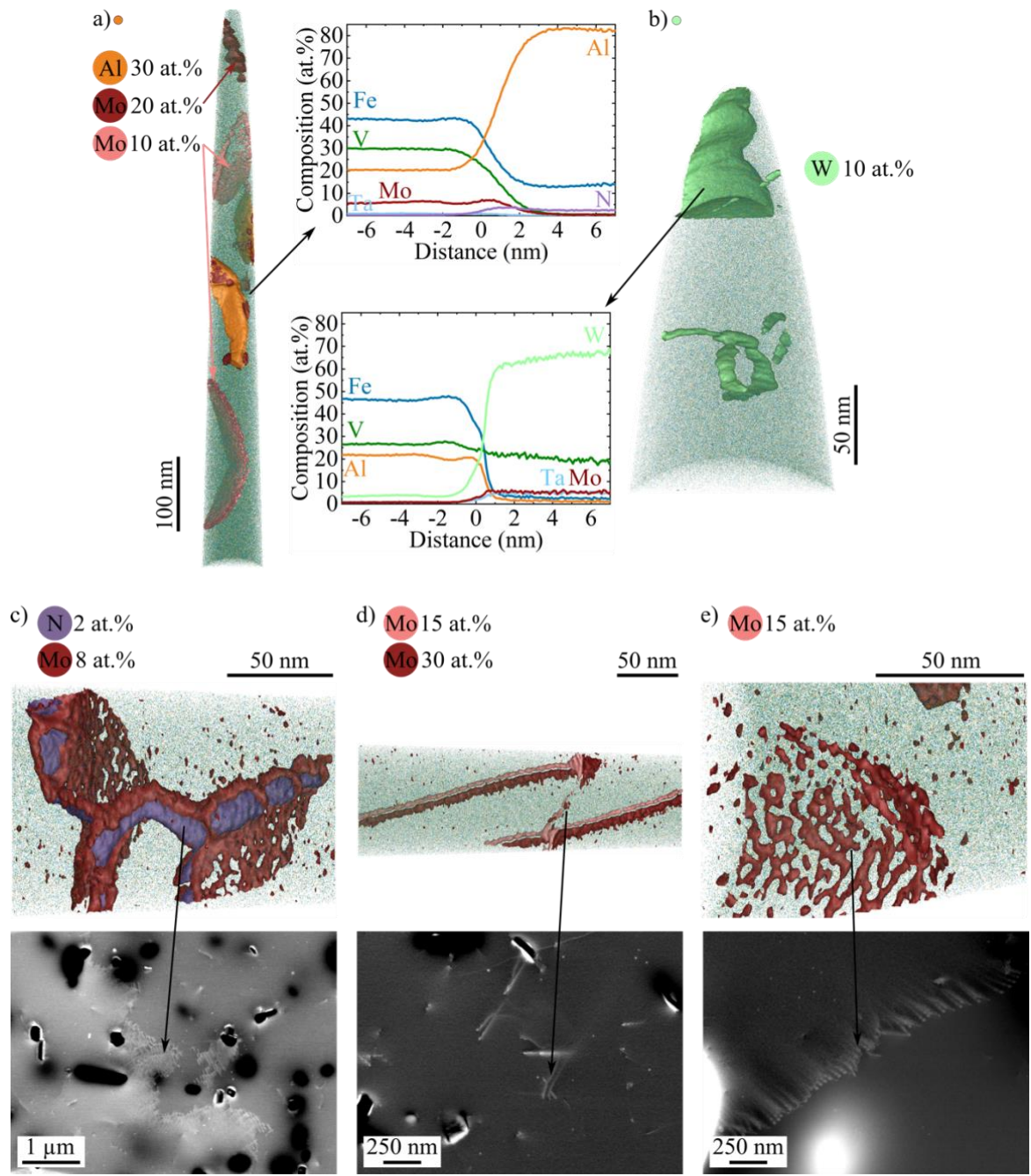

*Fig. 4* APT reconstructions of different regions within the SPS$_{5\,min}$ sample. a) Specimen taken from a region close to a high-angle grain boundary, b) specimen taken from a region with W precipitates. Proxigrams of the Al and W precipitates are shown, indicating the composition of the precipitates. c-e) Corresponding APT reconstructions and ECCI micrographs of typical observed features: c) entangled dislocations, most likely forming a LAGB, from which plate-shaped precipitates appear to form, d) a combination of partial dislocations Cottrell atmospheres, e) LAGB consisting of parallel dislocations.

### 3.3 Physical properties measurements

To relate microstructure to electron and phonon scattering, and charge carrier density, we used first a local *in-situ* four-point probe technique to obtain the electrical resistivity at room temperature [31]. Five grains of different compositions, as well as two GBs, are analyzed. Their composition, measured by EDX, is given in **table S2**. **Figure 5** a-c shows the results of the local measurements, including an electron micrograph of the respective grains and an example of the electrical measurement configuration. Results show that doping $Fe_2VAl$ with Ta, W or Mo decreases the electrical resistivity inside the grains despite increasing the scattering centers by solute atoms and precipitates, indicating a change in density of states at the Fermi level - due to doping. The actual composition only has a minor effect on the resistivity, i.e. the carrier density is only related to dopants in solution, and the observed precipitation suggests that each has reached solubility limit. The resistivity is also lower at GBs, which are often decorated with e.g. Al, Figure 3a, or Al-rich particles, and can be expected to have a higher conductivity.

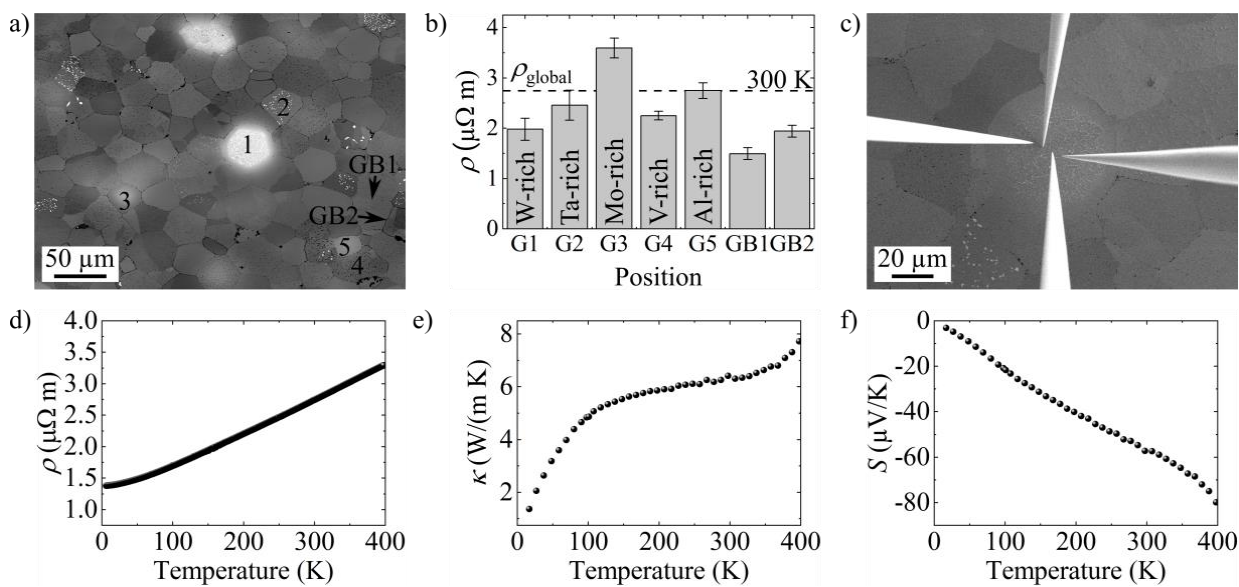

*Fig. 5* Measurement of the local electrical resistivity. 5 grains with different compositions and 2 grain boundaries were measured. a) SEM image showing the grains, b) Resistivity data, c) Local electrical resistivity measurement in-situ four-point probe technique, d) – f) bulk electrical resistivity, thermal conductivity, and Seebeck coefficient measurement results respectively.

In addition, the bulk electrical resistivity, thermal conductivity, and Seebeck coefficient were measured from 16 K to 400 K in the $SPS_{5min}$ sample. The results are summarized in figure 5 d-f. The electrical resistivity agrees with the local measurements, and shows an increase at higher temperatures, indicating a metallic behavior in this temperature range, as previously reported [32–34]. The local measurements suggest that this may be associated with the lower electrical resistivity at GBs compared to the core of the grains. The thermal conductivity is reduced to approx. 6.5 W/(m K), amongst the lowest values reported for this material [1,34,35]. Compared to the undoped, cast $Fe_2VAl$ [1], the thermal conductivity at room temperature is reduced by approx. 75%, while the electrical resistivity is reduced by approx. 63%, i.e. both favoring enhanced TE performance. However, the Seebeck coefficient is reduced compared to cast $Fe_2VTa_{0.05}Al_{0.95}$ [34], which can be related to the non-optimized charge carrier density arising from the high impurity level.

## 4   Summary and conclusion

To summarize, we show that designing a complex microstructure combined with doping by multiple elements, helps to increase the performance of $Fe_2VAl$-based TE materials, i.e. from a zT-value of 0.01 [1] to 0.05 at room temperature, and increasing to 0.1 at 400K. The complex microstructure of $Fe_2VTa_{0.05}Al_{0.95}$ powder was compared to bulk samples after SPS to reveal changes within the microstructure. The dendritic structure of the powder is lost due to sintering.

The microstructure of the compacted material certainly exhibits a hierarchy of property-enhancing microstructural features, including structural defects of different dimensionalities and precipitates, each with their respective composition. The limited resolution or probed volume of the analysis techniques employed make it difficult to reach a holistic analysis of all features across the extremely complex microstructure obtained after SPS. The unintended integration of W and Mo hinders the optimization of the charge carrier density, which lowers the Seebeck coefficient. Our results pave the way to enable the $Fe_2VAl$-based TE generators by using a combination of a defect-rich microstructure with an optimized charge carrier density to better control the Seebeck coefficient, and optimize grain boundary composition to further enhance electrical conductivity, offering further optimization of the TE performance.

**Acknowledgment**

L. G. gratefully acknowledges IMPRS-SurMat. U. Tezins, A. Sturm, C. Broß, M. Nellessen, and K. Angenendt are acknowledged for their technical support at the FIB/APT and SEM facilities at MPIE. L. G. and B.G. are grateful for financial support from the DFG on project GA 2450/4-1. H.B. acknowledges the financial support by the ERC Advanced Grant GB CORRELATE (Grant Agreement 787446 GB-CORRELATE) of Gerhard Dehm. B.G., K.P.S. and O.G. acknowledge also support from the Deutsche Forschungsgemeinschaft (DFG, German Research Foundation) by the CRC "HoMMage" Project-ID 405553726–TRR 270.

**Supporting Information**

Supplementary Information:

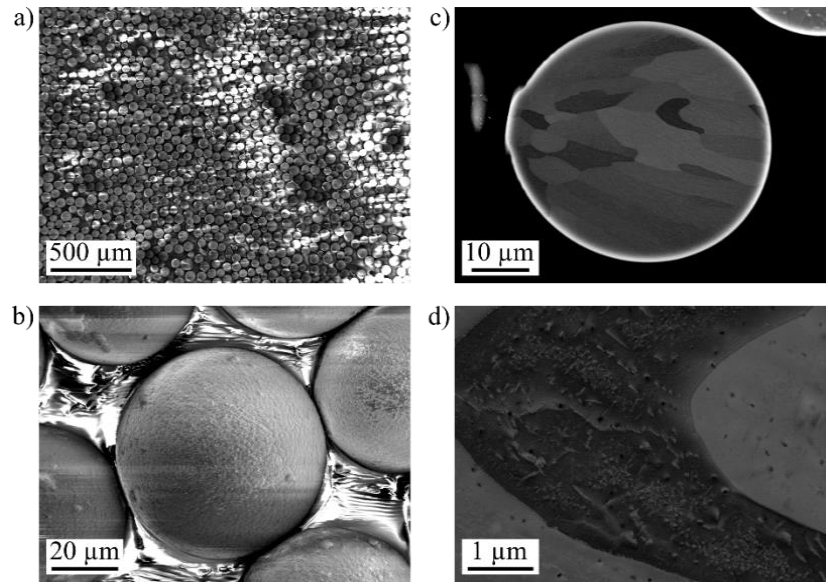

**Fig. S1**: a-b) SEM images in secondary electron mode of the as-received powder particles. These images have been taken with an acceleration voltage of 5 kV due to charging effects, which degrade the image quality, c-d) SEM images in backscattered electron mode of the cross section of a polycrystalline particle.

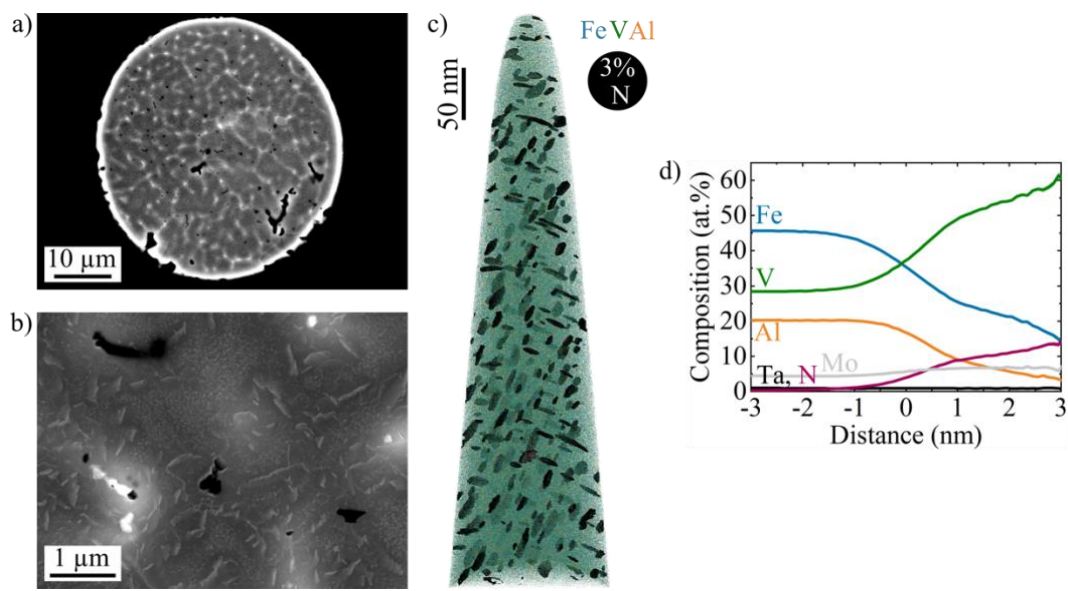

*Fig. S2: a) Backscattered electron micrographs of a $Fe_2VTa_{0.05}Al_{0.95}$ powder particle, which shows a dendritic structure with bright imaged interdendritic regions. b) Close-up of the microstructure, showing a high density of nano-precipitates. c) APT reconstruction, showing a high density of disc-shaped $VN_x$ precipitates, indicated by a black iso-composition surface of 3 at.% N. Fe, V, and Al are shown as blue, green, and orange dots, and d) proxigram of the precipitates.*

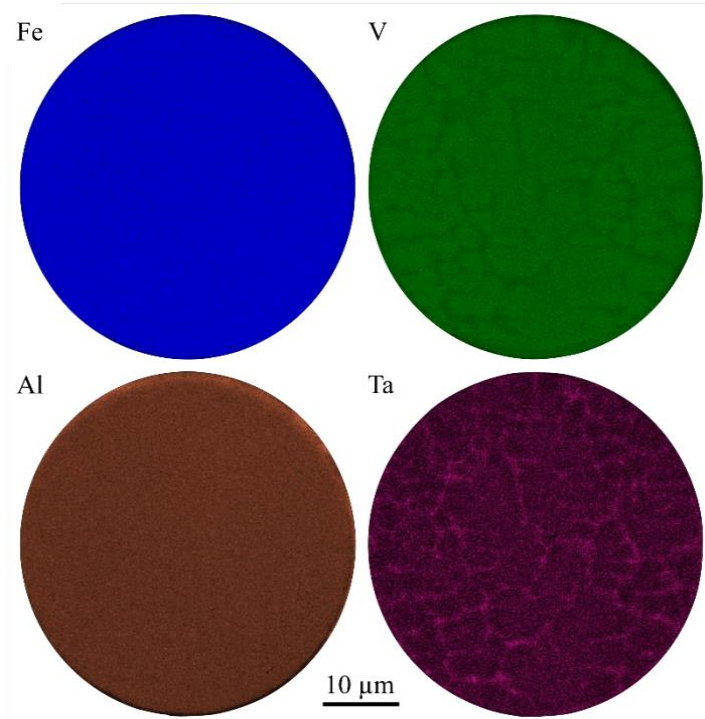

**Fig. S3**: EDX mapping of a dendritic powder particle. The regions imaged brightly in the BSE image in Fig. 1b is enriched in Ta and depleted in V. Fe and Al are homogeneously distributed on the scale of SEM-EDX measurements.

Table S1: Evaluation of the composition of different powder particles, based on 125 individual particles.

| Fraction | Fe | V | Al | Ta | Mo | W |
|---|---|---|---|---|---|---|
| 26 % | X | X | X | X | 0 | 0 |
| 3 % | X | X | X | - | 0 | 0 |
| 3 % | X | + | - | X | 0 | 0 |
| 31 % | X | X | X | X | <2 at.% | 0 |
| 9 % | X | X | - | X | <2 at.% | 0 |
| 2 % | X | X | X | - | <2 at.% | 0 |
| 7 % | X | + | - | X | <2 at.% | 0 |
| 1 % | X | + | X | + | <2 at.% | 0 |
| 2 % | X | X | X | X | 2 -5 at.% | 0 |
| 4 % | X | X | - | X | 2 -5 at.% | 0 |
| 2 % | X | X | X | X | 0 | <1 at.% |
| 7 % | X | X | X | X | <2 at.% | <1 at.% |
| 2 % | X | + | - | X | <2 at.% | <1 at.% |
| 1 % | - | + | - | X | 30 at.% | 0 |
| 1 % | X | X | - | X | 0 | 6 at.% |

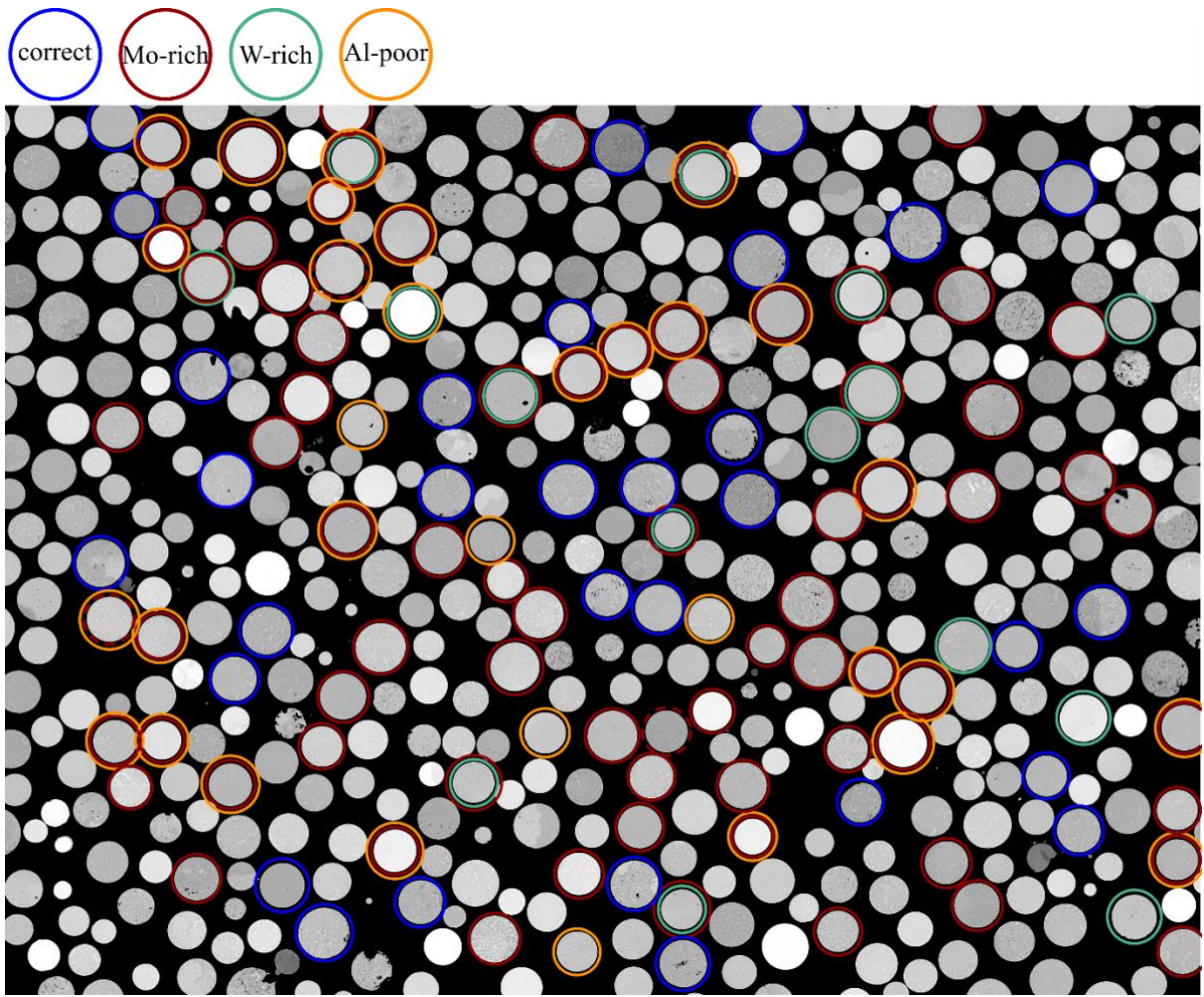

**Fig. S4**: BSE image of powder particles and respective composition, indicated by the circles. No correlation between composition and particle microstructure is found, yet, the Z-contrast is observed and particles with a high concentration of Mo and W appear brighter than particles with the nominal composition.

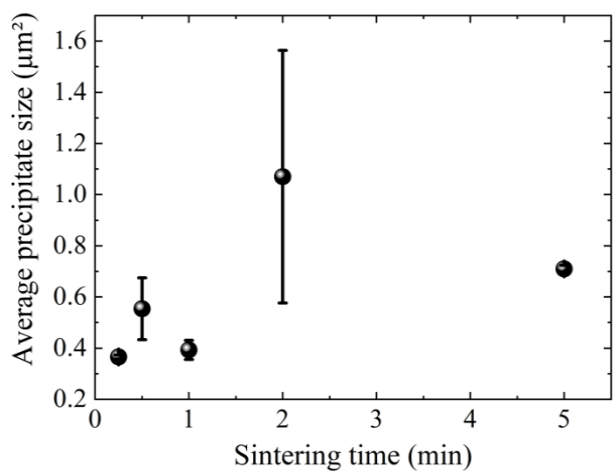

**Fig. S4**: Average size of Ta precipitates for different sintering times. Indications of precipitate growth is observed, albeit with low statistics and partly large errors.

**Tab. S2**: Composition of the grains used for the measurement of the local electrical resistivity, determined by EDX.

|    | Fe (at.%)    | V (at.%)     | Al (at.%)    | N (at.%) | Ta (at.%)    | Mo (at.%)    | W (at.%)     |
|----|--------------|--------------|--------------|----------|--------------|--------------|--------------|
| G1 | 45.5 ± 2.7   | 26.5 ± 2.9   | 20.3 ± 7.8   | NA       | NA           | NA           | 6.6 ± 4.0    |
| G2 | 43.8 ± 2.7   | 26.5 ± 3.0   | 19.0 ± 7.7   | NA       | 9.4 ± 3.3    | NA           | NA           |
| G3 | 42.8 ± 2.5   | 27.2 ± 2.7   | 21.8 ± 7.8   | NA       | NA           | 7.1 ± 5.4    | NA           |
| G4 | 23.9 ± 1.9   | 37.3 ± 1.2   | 10.9 ± 1.0   | NA       | 0.6 ± 0.5    | NA           | NA           |
| G5 | 14.0 ± 0.8   | 14.0 ± 0.8   | 14.0 ± 0.8   | NA       | 14.0 ± 0.8   | 14.0 ± 0.8   | 14.0 ± 0.8   |